\pdfoutput=1
\documentclass[aps,prl,twocolumn,a4paper,floatfix]{revtex4}
\usepackage[latin1]{inputenc}
\usepackage{graphicx}
\usepackage{amsmath,amssymb}
\usepackage{hyperref}
\usepackage{datetime}
\usepackage{pgfpages}
\usepackage{tikz}

\newcounter{lastnote}

\begin{document}

\title{Conduction of Ultracold Fermions Through a Mesoscopic Channel}

\author{Jean-Philippe Brantut} 
\affiliation{Institute for Quantum
Electronics, ETH Zurich, 8093 Zurich, Switzerland}
\author{Jakob Meineke} 
\affiliation{Institute for Quantum
Electronics, ETH Zurich, 8093 Zurich, Switzerland}
\author{David Stadler} 
\affiliation{Institute for Quantum
Electronics, ETH Zurich, 8093 Zurich, Switzerland}
\author{Sebastian Krinner}
\affiliation{Institute for Quantum
Electronics, ETH Zurich, 8093 Zurich, Switzerland}
\author{Tilman Esslinger} 
\email{esslinger@phys.ethz.ch}
\affiliation{Institute for Quantum
Electronics, ETH Zurich, 8093 Zurich, Switzerland}

\date{\pdfdate}

\begin{abstract}

In a mesoscopic conductor electric resistance is detected even if the device is defect-free.  We engineer and study a cold-atom analog of a mesoscopic conductor. It consists of a narrow channel connecting two macroscopic reservoirs of fermions that can be switched from ballistic to diffusive. We induce a current through the channel and find ohmic conduction, even for a ballistic channel. An analysis of {\it in-situ} density distributions shows that in the ballistic case the chemical potential drop occurs at the entrance and exit of the channel, revealing the presence of contact resistance. In contrast, a diffusive channel with disorder displays a chemical potential drop spread over the whole channel. Our approach opens the way towards quantum simulation of mesoscopic devices with quantum gases.

\end{abstract}

\maketitle

The quantum simulation of models from solid-state physics using cold atoms has seen tremendous progress over the last decade \cite{bloch_many-body_2008,esslinger_fermi-hubbard_2010}. Still, there are only limited analogies to the concept of conduction, which is at the core of mesoscopic solid-state physics. To close this gap it would be highly desirable to connect a probing region in a cold atom experiment to external incoherent reservoirs. This would lead to directed transport, the control of which is the basis of electronics. In such an intrinsically open configuration, boundary conditions play a crucial role, as in the Landauer theory of transport \cite{imry_introduction_1997}. So far, the transport properties in cold atom systems have been investigated by observing the response of the system to variations of the external potential \cite{jin_collective_1996,mewes_collective_1996,ben_dahan_bloch_1996,ott_collisionally_2004, billy_direct_2008,roati_anderson_2008,kondov_three-dimensional_2011,sommer_universal_2011,Schneider_fermionic_2012,Jendrzejewski:2012vn}, or by monitoring the coherent evolution of bimodal Bose-Einstein condensates\cite{ albiez_direct_2005, schumm_matter_2005, levy_the_2007} as a response to a bias. Extending the concept of quantum simulation to conduction requires the engineering of macroscopic reservoirs, an atom battery or capacitor connected to the conductor \cite{seaman_atomtronics_2007,das_quantum_2009,bruderer_mesoscopic_2012}. Such experiments would not only provide cold atoms with a faithful modeling of conduction experiments in solids, but also open the possibility to study phenomena that emerge at the contacts of mesoscopic systems \cite{imry_conductance_1999}.

We report on the observation of atomic conduction between two cold atom reservoirs through a mesoscopic, multimode channel. Our observations are three fold: (i) After preparing the two reservoirs with different atom numbers, we measure the atom current as a function of time. The system slowly evolves towards equilibrium, realizing a quasi-steady state at each point in time. This allows us to extract the current as a function of the atom number difference. (ii) Using {\it in-situ} imaging of the atomic density in the channel, we establish that the chemical potential drop occurs at the contacts between the channel and the reservoirs, as predicted by the Landauer theory. (iii) Alternatively, we prepare a strongly disordered channel having the same macroscopic conduction properties, and observe that the chemical potential drop is spread over the whole channel, revealing a finite linear resistivity. 

\begin{figure}[htb]
    \includegraphics[width=55mm]{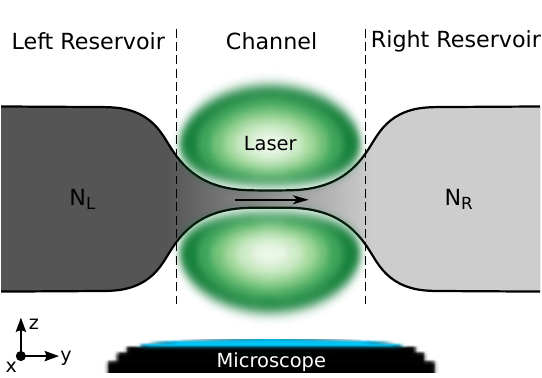}
    \caption{Experimental configuration. A macroscopic $^6$Li cloud is divided into two reservoirs separated by a narrow channel. The channel is imprinted using the two lobes in the intensity profiles of a nearly TEM$_{01}$-mode laser beam at the wavelength of 532 nm, created with a holographic plate. The distance between the two lobes is 18 $\mu$m, and the waist of the beam in the $y$-direction is 30 $\mu$m. A microscope objective (NA = 0.55) is used to observe or manipulate the atoms in the channel. When $N_L$ is larger than $N_R$, an atomic current $I$ flows through the channel (represented by the arrow on the figure).  }
    \label{fig:setup}
\end{figure}

Our measurement is made possible by the separation of scales in our trap geometry, as illustrated in figure \ref{fig:setup}. The experimental configuration consists of two identical, macroscopic cold atom reservoirs, which contain the majority of the atoms and feature fast equilibration dynamics. They are connected by a multimode channel, which contains a negligible fraction of the atoms, and supports a few quantum states in the $z$-direction, while it has the same extension as the reservoirs in the $x$-direction, making it quasi-two-dimensional. A high-resolution microscope is used to detect and influence the atoms at the micrometer scale \cite{zimmermann_high-resolution_2011}.

We prepare quantum degenerate gases containing $N_{\mathrm{tot}}$ = $4 \cdot 10^4$ $^6$Li atoms in each of the two lowest hyperfine states at a temperature of $0.36(18)$ $T_F$, where $T_F \simeq 700$\,nK is the Fermi temperature in a combined optical and magnetic trap \cite{zimmermann_high-resolution_2011}.
A laser beam propagating along the $x$-direction is focused on the center of the atomic cloud. The beam has a nodal line in the middle of its intensity profile and produces a repulsive potential for the atoms, which is tightly confining in the $z$-direction \cite{meyrath_high_2005,smith_quasi-2d_2005}. Oscillation frequencies of up to 3.9\,kHz along the $z$-direction are achieved, see figure 1.

\begin{figure}[htb]
    \includegraphics[width=55mm]{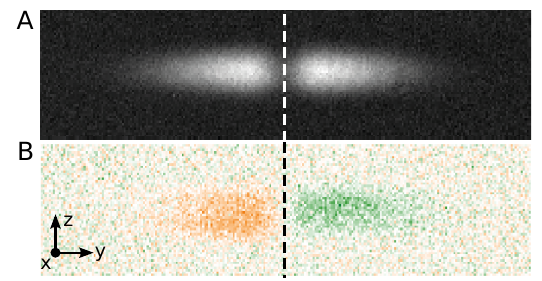}
    \caption{Atomic reservoirs connected by a mesoscopic channel. A : Absorption image of the atoms before the imbalance is applied. The image is taken after 1\,ms expansion in the $x$-$z$ plane. The dark region at the center reveals the presence of the mesoscopic channel. The dashed line tracks the position of the channel on both panels. B : Density difference between the cloud in the unbalanced configuration, as the current sets in, and the equilibrated cloud. Excess of atoms is displayed in orange, and lack of atoms is displayed in green. The imbalance of the reservoirs $\Delta N/N_{\mathrm{tot}}$ is set to 0.2. } 
   \label{fig:images}
\end{figure}

Figure \ref{fig:images}A presents a typical absorption picture of a cloud in the presence of the channel. We observe two clouds clearly separated by a low density region, revealing the presence of the channel and confirming that it contains a negligible fraction of the total atom number (smaller than 0.01). 

The conduction measurement proceeds as follows. We create an asymmetry in the potential by applying a constant magnetic field gradient of 2.5 $\rm mT\,m^{-1}$ along the $y$ axis. This is done during the evaporation process and eventually results in an imbalance $\Delta N/N_{\mathrm{tot}}$ $\sim$ $0.2$, where $\Delta N$ is the number difference between the two reservoirs. After evaporation, the confining potential of the trap is increased and a uniform magnetic field is set to 47.5\,mT. At this value, the scattering length of atoms in the two internal states is -100$\,a_0$, with $a_0$ being the Bohr radius. This ensures that the collision rate is sufficient to maintain thermal equilibrium in each reservoir on a time scale of $\simeq 30$\,ms. It also ensures that the mean free path ($\simeq1.3$\,mm) is much larger than the length of the channel, making it ballistic. The symmetry of the trapping potentials is then restored by switching off the magnetic field gradient in $50$\,ms, a time longer than the internal thermalization time of each reservoir, but short compared to the timescale of equilibration of the populations of the two reservoirs. Figure \ref{fig:images}B shows the difference between an absorption picture taken with and without imposing an imbalance. The left reservoir is seen to contain an excess of particles compared to the balanced reservoirs situation, and the right reservoir shows a deficit of particles. 

\begin{figure}[htb]
    \includegraphics[width=55mm]{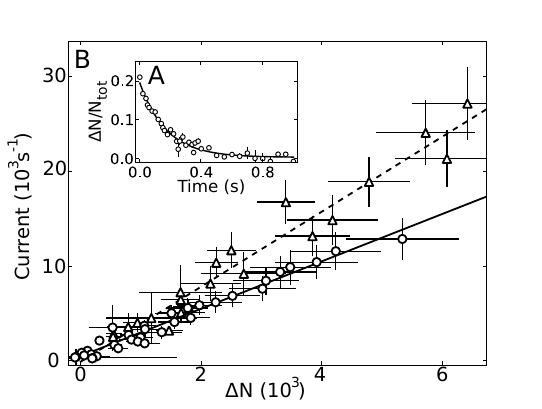}
    \caption{Observation of ohmic conduction. A: measured imbalance between the two reservoirs as a function of time. The solid line is an exponential fit to the data. B : Current as a function of number difference between the two reservoirs, measured from the exponential fit of figure A, for two different confinements in the channel. A small offset obtained from the fits in figure A, which is due to a slight misalignment of the channel with respect to the center of the trap, has been substracted. Circles : maximum center frequency along $z$ set to 3.9 kHz, triangles : 3.2 kHz. The lines are linear fits to the data. }
    \label{fig:discharge}
\end{figure}

The equilibrium of the whole system is characterized by a balanced population of both reservoirs, thus after restoring the symmetry of the trap an atomic current sets in through the channel. Figure \ref{fig:discharge}A presents the time evolution of $\Delta N/N_{\mathrm{tot}}$, with the oscillation frequency along $z$ in the channel set to $3.9$\,kHz. We observe an exponential decay (solid line on figure \ref{fig:discharge}A), with a time constant of $170$(14)\,ms. This exponential shape suggests a direct analogy with the discharge of a capacitor through a resistance. Indeed, the evolution of the system can be described as 
\begin{eqnarray}
\frac{d}{dt} \Delta N &=& -\frac{G}{C} \Delta N 
\label{eq}
\end{eqnarray}
where $G$ is the conductance of the channel, $C$ = $\frac{\partial N}{\partial \mu}$ is the compressibility of the reservoirs and $\mu$ is the chemical potential. The compressibility is analogous to the capacity of a capacitor. We neglect possible thermoelectric effects, since we do not observe a noticeable temperature evolution in the reservoirs.

Because the decay is the slowest process, the derivative of the curve around any point is a measurement of the current at a certain number difference, where the atoms in each reservoir have a known, thermal distribution. Therefore, the magnitude of the current measures the DC characteristic of the channel. Figure \ref{fig:discharge}B shows the observed current as function of the number difference for the same data set (circles), and for a channel with reduced confinement of 3.2\,kHz at the center (triangles). A linear relation is manifest for both cases, which confirms dissipative, ohmic conduction and allows to extract the slopes $G/C$ = 2.9(4)\,$s^{-1}$ and 3.7(2)\,$s^{-1}$ respectively. 

The observation of resistance in the ballistic conduction shows that the boundary conditions are essential in the investigation of transport, as in the Landauer approach. Indeed, the free expansion of a non-interacting cloud is also ballistic, but the absence of connection to reservoirs leads to the absence of any resistance to the flow, other than inertia. Furthermore this ballistic expansion generally does not depend on the conduction properties of the initial cloud, as ballistic expansion has even been observed for a band insulator \cite{Schneider_fermionic_2012}. 

The Landauer-B\"uttiker formula states that at zero temperature, the conductance of a ballistic conductor is equal to $1/h$ per quantum state contributing to the conduction, where $h$ is Planck's constant \cite{imry_conductance_1999}. Due to the quasi-two-dimensional character of the channel, the current is carried by many transverse modes, which are not individually resolved because of finite temperature. Instead, as the channel confinement is varied, the conductance is expected to vary linearly with the oscillation frequency along the confined direction $z$, due to the variations in the number of modes available. In both measurements, the reservoirs have the same capacities, thus the ratio $0.76(11)$ of the two slopes is equal to the ratio of conductances alone, and agrees qualitatively with the ratio $0.82$ of trap frequencies along $z$. We found that the linear relation between resistance and trap frequency persists for various confinement within the accessible range. The contact resistance, which naturally appears in the Landauer picture, explains the observation of Ohmic conduction even in our defect free channel. While every atom that enters the channel on one side exits on the other with the same momentum with probability one, only a tiny fraction of the atoms from each reservoir can pass through the channel at any given time due to the Pauli principle. The atoms incident on the channel therefore have to redistribute, leading to a drop of chemical potential in the contact region, while the chemical potential stays constant over the channel \cite{gao_four_2005,datta_electronic_1995}.

\begin{figure*}[htb]
    \includegraphics[width=125mm]{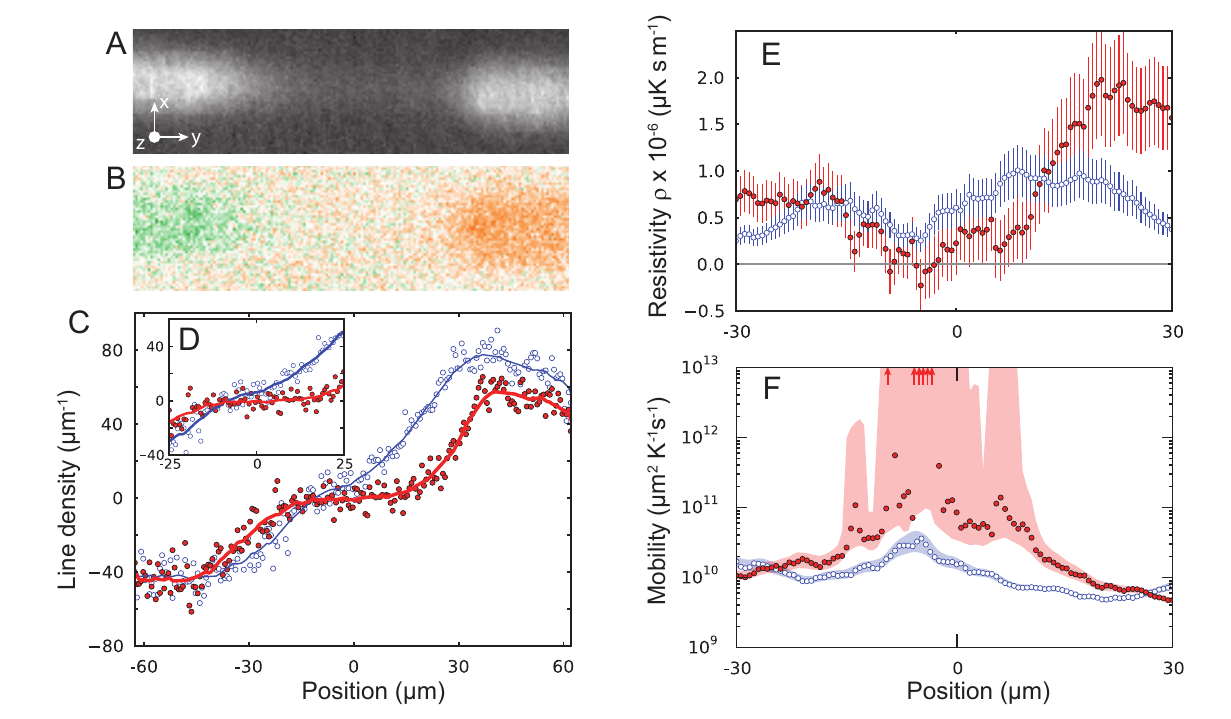}
    \caption{Investigation of conduction using high-resolution microscopy. A: Absorption picture of the density in the channel, for a cloud at equilibrium (no current). B: Difference between two pictures taken at equilibrium and with a current of $10^4\,s^{-1}$. The color is orange for positive difference and green for negative. C: Line-density difference obtained by accumulating B along the $x$-axis, for a ballistic channel (red full circles) and for a diffusive channel having the same conductance (blue open circles). The solid lines are smoothed data to guide the eye. D: Focus on the central part of the line density difference. E: Resistivity per unit length and time in the channel, as a function of position, for the ballistic (red full circles) and diffusive channel (blue open circles), computed from equation \eqref{diss}. Only the center of the channel is shown, where the extraction procedure for the line-compressibility is valid (see Materials and Methods for the details). The solid line indicates zero resistivity. F : Mobility in the channel as a function of position for the ballistic (red full circles) and diffusive (blue open circles) channels. The red arrows indicate points where the mobility is measured to be infinite. The error bars and shaded region reflect the uncorrelated combination of estimated systematic and statistical uncertainties.}
    \label{fig:insitu}
\end{figure*}

To gain further insight into this mechanism, we use high-resolution microscopy to observe the density distribution of atoms in the channel. We do so using {\it in-situ} absorption imaging along the $z$-direction, with and without current flowing through the channel. A typical picture of the density distribution in the channel in the absence of current is presented on figure \ref{fig:insitu}A. At the sides of the picture we observe the contacts with the two reservoirs which extend beyond the field of view. Closer to the center, the lower column density reveals the presence of the channel, which is smoothly connected to the reservoirs. Figure \ref{fig:insitu}B shows the difference between two such pictures, taken with and without current flowing through the channel. We see the small density difference between the two reservoirs, which reflects the macroscopic number difference shown in figure \ref{fig:images}B. 

The red points in figure \ref{fig:insitu}C show the line-density difference $\tilde{n}_{l}$ along the channel, obtained by accumulating image \ref{fig:insitu}B along the $x$-direction. At the center of the channel, the difference is close to zero over $30$\,$\mu$m, while the density difference changes quickly at the sides of the channel. This qualitative difference between the channel and the contacts indicates a localized chemical potential drop, and thus the presence of contact resistance \cite{trans_individual_1997}. Indeed, following the approach of mesoscopic physics \cite{datta_electronic_1995}, we introduce a local chemical potential $\mu(y)$, by requiring that it yields the observed density when used in the Fermi-Dirac distribution. As the mean free path for atomic collisions is very large, the energy distribution of atoms may not be thermal outside the reservoirs. Therefore, the definition of $\mu$ does not correspond to the local density approximation, and does not suppose local equilibrium. The variation of the local chemical potential is the counterpart of the electric field created by charge density variations along an out-of-equilibrium, ballistic electric conductor \cite{ulreich_where_1998}. 

From the current $I$ measured across the channel, we deduce from figure \ref{fig:insitu}C the local resistivity \cite{materialsandmethods}
\begin{eqnarray}
\rho = \frac{1}{\kappa I}\frac{\partial \tilde{n}_{l}}{\partial y} , \label{diss}
\end{eqnarray}
where we have introduced the line-compressibility in equilibrium $\kappa_l$ = $\partial n_{l} / \partial \mu$, with $n_{l}$ being the line-density along the $y$-axis. Figure \ref{fig:insitu}E presents (in red) the resistivity obtained by applying equation \eqref{diss} to the {\it in-situ} picture. The line-compressibility is obtained directly from the column density at equilibrium and the shape of the trap \cite{muller_local_2010,materialsandmethods}. At the center of the channel, the local resistivity is zero within error bars, whereas it quickly increases at the contacts between the channel and the reservoirs. We attribute the asymmetry between left and right to deviations from an ideal gaussian shape of the trapping beam away from the center. 

As the mean free path for atomic collisions is much larger than the channel, the resistive current observed does not lead to local heating. The energy associated with the current is released deep inside the reservoirs. The local resistivity observed rather indicates variations of momentum, which happen at the length scale of the variations of the external potential \cite{datta_electronic_1995}, leading to a local increase of entropy due to the loss of the information on the direction of the momentum. 

Many quantities of interest can be extracted from the microscopic density distribution. For instance, the drift velocity $v_d$ = $I/n_{1D}$ at the center of the channel is found to be $200\,\mu {\rm m}$\,s$^{-1}$, or $4\cdot10^{-3}$\,$v_F$, where $v_F$ is the Fermi velocity in the reservoirs, which confirms that our system realizes the Laudauer paradigm of conduction. We also introduce an atomic mobility for the atoms $v_d \kappa \left( \frac{\partial n_{1D}}{\partial y} \right)^{-1}$, which relates the drift velocity to the chemical potential gradient and thus characterizes the intrinsic conduction properties of the channel, regardless of the density. Figure \ref{fig:insitu}F presents the atomic mobility as obtained from the {\it in-situ} pictures for the ballistic channel (in red). We observe more than one order of magnitude increase of the mobility at the center of the channel, where obtained mobilities diverge, consistent with the expected infinite mobility of a purely ballistic channel. 

As opposed to the ballistic channel, we have engineered a channel where the conduction is diffusive, which is the case encountered in typical solid materials. To do so, we project a blue-detuned laser speckle pattern onto the channel, realizing a quasi-2D disorder \cite{robert_anisotropic_2010}. This pattern has a gaussian envelope with an rms diameter of $32$\,$\mu$m, an average amplitude of $0.6$\,$\mu$K at the center and a correlation radius of $0.37$\,$\mu$m \cite{materialsandmethods}. We then reduce the confinement of the channel down to $1.6$\,kHz along $z$, so that the atomic conductance of the disordered channel is the same as that of the ballistic one studied before. We thus have a second system displaying the same macroscopic transport properties, but with a different conduction mechanism. The measured line-density difference in the disordered channel is shown in blue in figure \ref{fig:insitu}C. We observe a continuous decrease from right to left, indicating a chemical potential drop spread over the whole channel, even at the center of the channel as shown in \ref{fig:insitu}D. The calculated resistivity, shown in blue in figure \ref{fig:insitu}E shows moderate variations along the channel, in contrast to the ballistic case, and remains sizable at the center. The mobility of the disordered channel (in blue in figure \ref{fig:insitu}F) is much smaller in the channel than for the ballistic case, and clearly saturates to a finite value at the center.  

Our configuration is closely analogous to that of a field-effect transistor. The strength of the confinement in the channel has been used to vary the conductance by changing the density. Further tuning could be obtained by adding a repulsive gate laser, tuning the channel to fully isolating while keeping an infinite mobility. In addition, the effects of disorder in such a device can be studied systematically by varying the laser-induced random potential. Metal-insulator transitions, such as two-dimensional Anderson localization \cite{kuhn_localization_2005}, can be studied in a way which is directly analogous to real solid-state devices \cite{abrahams_metallic_2001}. The ability to further control the disorder could be used to study universal conductance fluctuations \cite{imry_introduction_1997}.  Apart from disorder, various potentials can be designed and projected onto the channel using the microscope setup \cite{zimmermann_high-resolution_2011}. This will allow us to measure the conduction properties of various model systems. For example, quantized conduction can be investigated if a single mode can be resolved in the channel \cite{Wharam_1988,van_wees_quantized_1988,thywissen_quantum_1999}. Furthermore, conductance is very sensitive to interactions between atoms, and would be an ideal observable to investigate strongly correlated fermions. The combination of mesoscopic atomic devices with controlled interactions opens fascinating perspectives and could shine new light on open questions in the field of mesoscopic physics \cite{yacoby_nonuniversal_1996}.

We acknowledge fruitful discussions with Gianni Blatter, Antoine Georges, Henning Moritz and Wilhelm Zwerger, and the help of Torben M\"uller during the early stage of the experiment. We acknowledge financing from NCCR MaNEP and QSIT, ERC project SQMS, FP7 project NAME-QUAM and ETHZ. JPB acknowledges support from EU through Marie Curie Fellowship.

\bibliographystyle{science}
\bibliography{paper}

\section*{Materials and Methods}

\subsection{Potential drop}

Consider a small region of the channel around position $y$. The line density around this point at equilibrium is $n_l(y)$ and the line density in the presence of current is $n_l(y)+\tilde{n_l}(y)$. 

Since there is no interparticle scattering at the length scale of that small region, the energy distribution in the presence of a current is not given by the equilibrium (Fermi-Dirac) distribution. Indeed, the density varies at the short scale of the scattering with impurities or with the external potential, while the distribution relaxes towards equilibrium on the scale of the interparticle interaction mean free path. 

Following the standard procedure in mesoscopic physics (see \cite{datta_electronic_1995}, chap.2 part 3), we attribute to the region around $y$ a chemical potential $\mu(y)$, which gives the expected density when inserted in the Fermi distribution, and coincides with the chemical potential in the reservoirs. In this way, the chemical potential variations are local (like the density variations), even if the energy relaxation takes place on a larger length scale, deep in the reservoirs. 

We now consider the relation between $\mu$, $\tilde{n_l}$ and the compressibility $\kappa_l = \partial n_l / \partial \mu$. Let $f(\mu,T,E)$ be the Fermi-Dirac distribution at temperature $T$, chemical potential $\mu$ and energy $E$, and $g(E)$ be the density of states, 

\begin{eqnarray}
\tilde{n_l} &=& \int f(\mu,T,E) -f(\mu_0,T,E) g(E) dE \\
 &=& (\mu-\mu_0) \frac{\partial}{\partial \mu} \int  f(\mu_0,T,E) g(E) dE \\
 &=& (\mu-\mu_0) \frac{\partial n_l}{\partial \mu} \\
 &=& \tilde{\mu} \kappa_l, \label{eq:kappa}
\end{eqnarray}
where we have introduced the local chemical potential at equilibrium $\mu_0$, which follows from the local density approximation at equilibrium, and $\tilde{\mu} = \mu-\mu_0$ the local deviation of the chemical potential away from equilibrium.

Knowing the current $I$ flowing through the channel, we now introduce a local resistivity 
\begin{equation}
\rho(y) = \frac{\partial \tilde{\mu}}{\partial y} \frac{1}{I}, 
\end{equation}
which reflects the local scattering with obstacles. Although the Joule heating happens deep in the reservoirs where excited atoms release their energy via collisions with other atoms, the momentum randomization due to scattering with impurities is local, and leads to a local increase of entropy as information on the direction of motion of particles gets lost. 

Experimentally, to get the derivative $\partial \tilde{n}_l/\partial y$ we fit a line to the line density difference at each position $y$ within a window of $\pm 9 \, \mu$m around this point. 

To extract $\kappa_l$ at position $y$ we make use of the isotropy
of the compressibility. We consider at fixed position $y$
the variation of $n_{col}$ along the $x$ direction and compare it with
the known trap shape \cite{muller_local_2010}.
Consider the gas around position $y$:
\begin{eqnarray}
\kappa_l(y) &=& \frac{\partial n_l(y)}{\partial \mu}\\
 &=& \frac{\partial}{\partial \mu} \iint n(\mu(x,y,z)) dx dz\\
  &=& \iint \frac{\partial n}{\partial \mu}(x,y,z)  dx dz\\
  &=& \iint \frac{\partial n}{\partial x} \left( \frac{\partial \mu}{\partial x} \right)^{-1} dx dz\\
  &=& -\iint \frac{\partial n}{\partial x} \left( \frac{\partial V}{\partial x} \right)^{-1} dx dz
\end{eqnarray}
The confinement along the $x$ direction is ensured by the optical dipole trap, and is constant over the channel.
Close to the center of the channel, the confinement along $z$ is mainly ensured by the two repulsive lobes of the laser beam. Therefore, the variations of the potential along the $z$ direction is independent of the variations along the $x$ direction and can therefore be pulled out of the $z$ integration.
\begin{eqnarray}
\kappa_l(y) &=& -\int  \left( \frac{\partial V}{\partial x} \right)^{-1} \left( \int \frac{\partial n}{\partial x} dz  \right)dx \\
 &=& -\int  \left( \frac{\partial V}{\partial x} \right)^{-1} \frac{\partial n_{col}}{\partial x} dx \label{kappa} \label{kappa_final} \, ,
\end{eqnarray}

where $n_{col}$ is the column density. 

A slice of $\pm 9 \, \mu$m around one position $y$ is taken from
the density picture at equilibrium (figure 4A) and the average along the $y$ direction is calculated.
This results in a one-dimensional density profile along the $x$ direction.
We fit a gaussian to such a profile and use it in equation \eqref{kappa_final},
in order to avoid the noise generated by numerical differentiations and 
ratios of those \cite{muller_local_2010}.
This fits the shape of the cloud within the errorbars. In
equation \eqref{kappa_final}, $V$ is taken to be the known gaussian
shaped trapping potential.

\subsection{Mobility}

In electric conduction the mobility relates the drift velocity $v_d$
to a potential gradient (electric field). In our case the gradient in
the chemical potential plays the role of the potential gradient
and the atomic mobility is 
\begin{eqnarray}
v_d \cdot \left(\frac{\partial \mu}{\partial y}\right)^{-1} = v_d \cdot \kappa_l \left(\frac{\partial n_l}{\partial y}\right)^{-1},
\end{eqnarray}
where we used equation \eqref{eq:kappa}.

In the data analysis we attribute an infinite atomic mobility
to points where $\frac{\partial \tilde{n}_l}{\partial y}$ is zero
or slightly negative. These points are indicated
with red arrows in figure 4. The upper limit of the red shaded region
is as well set to infinity if the lower edge of the error on
$\frac{\partial \tilde{n}_l}{\partial y}$ extents to negative values.

\subsection{Properties of the confining potential}

The channel is imprinted on the atoms using a 532\,nm wavelength laser beam, with waists $30.2(3)$ and $10.3(3)$\,$\mu$m along the $y$ and $z$ directions, respectively. This beam passes through a holographic plate (Silios) dephasing the upper part of the beam by $\pi$ with respect to the lower part \cite{smith_quasi-2d_2005,meyrath_high_2005}. The beam has been characterized in a test setup and shows a slight asymmetry along the $y$ axis away from the center. The contrast of the central region of the beam compared to the lobes is larger than $0.99$. In the parameter regime explored in the paper, the residual light leads to a repulsive potential along the $y$-axis smaller than $40$\,nK, much smaller than the oscillation frequency along $z$.

The oscillation frequency along the tightly confining direction $z$ has been measured {\it in-situ} using parametric heating on a microscopic cloud \cite{zimmermann_high-resolution_2011}, and the measured frequency agrees with the fitted curvature of the intensity profile observed on the test setup. Along the propagation direction of the beam, the measured curvature of the intensity profile varies by less than 5\% on a length scale of 200\,$\mu$m, and is therefore constant over the transverse radius of the cloud. 

\begin{figure*}[htb]
\def \thefigure{S1}
    \includegraphics[width=125mm]{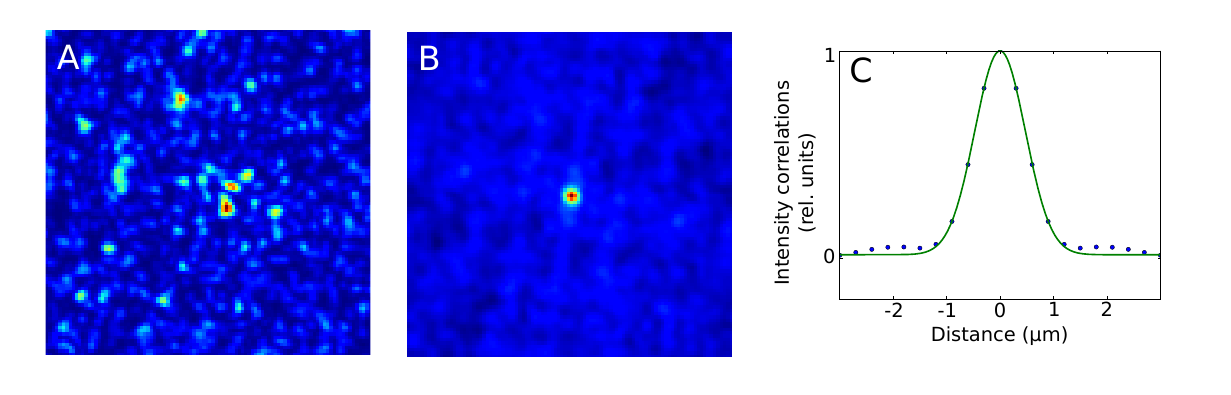}
    \caption{Properties of the disordered potential. A : central region of the disordered potential as observed with the high-resolution microscope used to image the atoms. The total width of the picture is $30$ $\mathrm{\mu}$m. B : correlation function of the intensity distribution observed in A. C : horizontal cut of B, showing the correlation properties of the disordered potential (in blue). The solid green line shows a Gaussian fit yielding a measured correlation radius of $460$\,nm.}
    \label{fig:speckle}
\end{figure*}

\subsection{Properties of the disordered potential}

The disordered potential is created by passing through a light diffuser (Luminit) with a beam at 532 nm, then through a high resolution microscope objective \cite{zimmermann_high-resolution_2011}, corrected for aberrations at that wavelength. The resulting disordered potential is observed directly {\it in-situ} using a second microscope identical to the first. In this way, we directly characterize the potential correlation properties by taking pictures of the potential as seen by the atoms in the glass cell. Figure \ref{fig:speckle} presents a typical observation of the potential characteristics. Figure \ref{fig:speckle}A shows a zoom on the central part of the disordered potential where correlation properties are computed. Figure \ref{fig:speckle}B shows the measured correlation function obtained using the inverse Fourier transform of the power spectrum of A. Figure \ref{fig:speckle}C shows a cut along the horizontal axis, of figure B, together with a Gaussian fit, yielding an observed correlation radius (1/$\sqrt{e}$ radius) in this direction of $460$\,nm. In the other direction, the same fit yields an observed correlation radius of $560$\,nm. To obtain a faithful estimate of the correlation properties of the disorder on the atoms, we deconvolve the correlation properties using the measured, gaussian point spread function of the microscope used to image the pattern \cite{zimmermann_high-resolution_2011}. This yields a correlation radius averaged for the two directions of $370$\,nm. In addition, using the {\it in-situ} images of the light intensity, we also observed the expected exponential probability distribution of intensities. 

The envelope of the disorder potential is straightforwardly obtained from a gaussian fit of the observed profile. The total power sent through the microscope, together with the average fitted profile of the envelope, yields the average depth of the disordered potential cited in the text.

\end{document}